\documentclass{article}
\usepackage{spconf,amsmath,graphicx}
\usepackage{xcolor}
\usepackage{amssymb}

\usepackage{caption,tabularx,colortbl}
\usepackage{graphicx}
\usepackage{booktabs}
\usepackage{comment}
\usepackage{tabularx}
\usepackage{bm}
\usepackage{array}
\usepackage{algpseudocode}
\usepackage{amssymb}
\usepackage{hyperref}

\newcommand{\X}{\mathbf{X}}
\newcommand{\x}{\mathbf{x}}

\newcommand{\e}{\mathbf{e}}
\newcommand{\B}{\mathbf{B}}
\newcommand{\A}{\mathbf{A}}
\newcommand{\W}{\mathbf{W}}

\newcommand{\I}{\mathbf{I}}

\newcommand{\y}{\mathbf{y}}
\renewcommand{\P}{\mathbf{P}}

\newcommand{\R}{\mathbb{R}}

\AtBeginDocument{
    \setlength{\abovedisplayskip}{4pt}
    \setlength{\belowdisplayskip}{4pt}
    \typeout{ABOVE DISPLAY = \the\abovedisplayskip}
    \typeout{BELOW DISPLAY = \the\belowdisplayskip}
    \typeout{ABOVE SHORT = \the\abovedisplayshortskip}
    \typeout{BELOW SHORT = \the\belowdisplayshortskip}

}

\AtBeginDocument{
}

\title{Sample Dependence-Aware Blind Source Separation \\ for Linear Causal Discovery}
\name{Cihan Eralp Kumbasar, Emin Erdem Kumbasar, Tülay Adali}
\address{\fontsize{11}{11}\selectfont Dept. of CSEE, University of Maryland Baltimore County, Baltimore, MD 21250}
\begin{document}
%
\maketitle
\begingroup
\renewcommand\thefootnote{}
\footnotetext{This work has been submitted to the IEEE for possible publication. Copyright may be transferred without notice, after which this version may no longer be accessible.}
\addtocounter{footnote}{-1}
\endgroup
\begin{abstract}
Causal discovery seeks to determine directional relationships among variables in complex systems, yet observational data alone generally do not identify causal direction. Complete identification therefore requires assumptions beyond linearity and acyclicity. A key approach is the Linear Non-Gaussian Acyclic Model (LiNGAM), which formulates linear structural equation model estimation as a blind source separation (BSS) problem. LiNGAM treats disturbances as mutually independent random variables under independent sampling, relying on higher-order statistics (HOS) and non-Gaussianity for identifiability. This restricts identifiability when multiple disturbances are Gaussian and ignores potentially informative temporal dependence. To broaden causal identifiability, we formulate linear acyclic causal discovery with mutually independent disturbance processes and transfer established BSS identification conditions to causal structure recovery. Under this formulation, replacing the mutual information cost with mutual information rate enables identification under a broader class of disturbances, including Gaussian processes with nonproportional covariance functions. Synthetic experiments reveal the limitations of methods relying solely on HOS or sample dependence, while their joint exploitation provides near-perfect causal recovery across the tested disturbance settings. Experiments on real functional magnetic resonance imaging (fMRI) data further demonstrate that jointly exploiting HOS and sample dependence improves the bootstrap stability of estimated ancestral directionality. 

\end{abstract}
\section{Introduction}
\label{sec:intro}

Understanding how variables causally influence one another is important for explaining complex systems. Structural equation models (SEMs) provide a principled framework for representing such relationships. However, in linear SEMs with Gaussian disturbances, observational data provide information primarily through the covariance structure, which is generally insufficient to uniquely identify the underlying directed causal graph \cite{pearl2000causality, spirtes2001causation}. 
Consequently, such methods typically recover a Markov equivalence class of structures and require additional assumptions or prior information, such as the causal order, to uniquely determine the causal structure \cite{lingam}. 
Linear Non-Gaussian Acyclic Model (LiNGAM) takes an alternative approach by exploiting non-Gaussianity to uniquely identify the linear causal structure from observational data, without requiring prior knowledge of the causal ordering \cite{lingam, directlingam}.

LiNGAM uses independent component analysis (ICA) \cite{bssbook} to solve the linear acyclic model and estimate the causal dependencies between variables. 
ICA provides a principled framework to blind source separation (BSS), with algorithms exploiting different forms of statistical diversity \cite{bssbook, icaivadiversityoverview}. 
Early source separation methods such as FastICA \cite{fastica} commonly rely on higher-order statistics (HOS) and make use of the non-Gaussianity of the source distributions. 
Even though methods that only use non-Gaussianity cover the majority of ICA algorithms, source separation strategies are not limited to the adoption of HOS as the diversity among source signals. Alternative types of diversities, such as sample dependence, can also be used for source separation \cite{icaivadiversityoverview, sobi, wasobi, erbm, erbmbss}. 

Classical LiNGAM models the disturbances as random variables sampled independently across observations, restricting the available BSS diversity to HOS and requiring that at most one disturbance is Gaussian for identifiability \cite{lingam,icaivadiversityoverview}. However, many real-world signals exhibit sample dependence, whose temporal structure can provide additional statistical diversity for identification \cite{icaivadiversityoverview,hyvarinenmeg}. Recent work has exploited this second-order temporal information for pairwise instantaneous causal discovery \cite{hyvarinenmeg}. A complementary direction is to jointly exploit HOS and sample dependence within a multivariate BSS formulation and transfer broader BSS identification conditions to causal structure recovery. 

In this work, we pursue this direction by extending the BSS interpretation underlying LiNGAM to mutually independent disturbance processes that may exhibit sample dependence. We transfer dependence-aware BSS identifiability results to linear acyclic causal modeling and show that combining sample dependence with HOS enables causal structure identification with Gaussian disturbance processes having distinguishable covariance functions. 
Synthetic experiments demonstrate the complementary roles of the two diversities: BSS methods relying solely on HOS fail with Gaussian disturbance processes, while methods relying only on sample dependence suffer when temporal diversity is insufficient. Joint exploitation of both achieves near-perfect causal recovery across diverse disturbance settings. 
Beyond controlled simulations, we apply causal discovery to real functional magnetic resonance imaging (fMRI) data and assess the stability of the inferred group causal structure through bootstrap resampling of subjects. 
Joint use of HOS and sample dependence yields more reproducible causal dependencies across variable pairs, while HOS-only methods show greater uncertainty in the directionality of ancestral relationships. 

\section{BSS-based Identification of Linear Acyclic SEMs}
\subsection{Linear Acyclic SEM as a BSS Model} \label{sec:linearsem}
The linear causal model assumes a linear, recursive process with acyclic causal relationships, allowing the variables to be ordered such that no later variable influences an earlier one \cite{lingam}.
Assuming each observed variable $x_i$ is zero mean, the relationship between observations is modeled as follows:
\begin{equation} \label{eq:lingam_sum}
    x_i = \sum_{k(j) < k(i)} {b_{ij}x_j + e_i},
\end{equation}
where $e_i$ is an external disturbance, $b_{ij}$ represents the causal influence strength from $x_j$ to $x_i$ and $k(i)$ denotes the causal order of $x_i$ in the DAG \cite{lingam}. We assume causal sufficiency (no unobserved common causes) and mutually independent disturbances with nonzero variances \cite{pearl2000causality, spirtes2001causation, lingam}.
The model can be rewritten in matrix form as:
\begin{equation} \label{eq:sem}
    \x  = \B\x + \e,
\end{equation}
where $\x \in \R^N$ is the random vector of observed variables, $\B \in \R^{N\times N}$ contains the causal coefficients $b_{ij}$, and $\e \in \R^N$ is the vector of external disturbances. The acyclic nature of causal relationships ensures that $\B$ can be permuted to a strictly lower triangular form. One can solve for $\x$ to write:
\begin{equation} \label{eq:lingamica}
    \x = \A\e,
\end{equation}
where $\A = (\I - \B)^{-1}$. This formulation defines an ICA model in which the disturbances $e_i$ are independent sources and $\A$ is the mixing matrix. The BSS algorithm estimates the corresponding demixing matrix $\W = \A^{-1} = \I - \B$.

The demixing matrix is identifiable only up to permutation and scaling. Under acyclicity, we permute its rows to align each estimated disturbance with its corresponding variable and normalize the diagonal to obtain \(\widehat{\W}\). We then estimate \(\widehat{\B}=\I-\widehat{\W}\) and determine a causal order by finding \(\P\) such that \(\P\widehat{\B}\P^\top\) is as close as possible to strictly lower triangular. Given this order, adaptive Lasso selects parents and ordinary least squares refits their coefficients. This procedure is shared by all BSS-based methods considered here \cite{lingam}. 

\vspace{-2mm}
\subsection{Identification Under Independent Sampling}
For a candidate nonsingular demixing matrix ${\W}$, let $\mathbf y={\W}\mathbf x$. 
Under independent sampling, ICA minimizes the mutual information among these source estimates $\y$:
\begin{equation} \label{eq:mutinf}
    \min_{\W} \mathcal{I}(\W)
    = \sum_{i=1}^N H(y_i) - H(\mathbf{y}),
\end{equation}
where $H(y_i)$ and $H(\mathbf{y})$ denote the marginal and joint entropies of the disturbance estimates, respectively. 
Through MI, ICA exploits non-Gaussianity through HOS and is identifiable when at most one disturbance is Gaussian, a condition inherited by ICA-LiNGAM for causal structure identification.

\vspace{-2mm}
\subsection{Random Process Disturbance Model}
We extend the disturbance model from mutually independent random variables to mutually independent random processes, whose samples are not required to be independent over time, while retaining the same linear acyclic causal structure:
\begin{equation}
    \x(t)=\B\x(t)+\e(t), \qquad
    \x(t)=\A\e(t). 
\end{equation}
The disturbances may therefore exhibit sample dependence, while temporally i.i.d. disturbances remain included as a special case. This allows temporal structure to provide an additional source of statistical diversity alongside HOS. 

\vspace{-3mm}
\subsection{Sample Dependence-Aware Identification}
The instantaneous MI objective in (4) does not exploit dependence across samples. 
For stationary random process disturbances with finite differential entropy rate, we use the mutual information rate (MIR), which jointly incorporates HOS and sample dependence: 
\begin{equation} \label{eq:mutinfrate}
    \min_{\W} \mathcal{I}_r(\W) = \sum_{i=1}^N {H_r(y_i)} - H_r(\mathbf{y}), \quad \y(t) = \W \x(t),
\end{equation}
where $H_r(y_i) = \lim_{t \rightarrow \infty} H[y_i(1), y_i(2), ..., y_i(t)]/t$ is the entropy rate of the $i$th source estimate, which measures the per-sample density of the average uncertainty of a random process \cite{icaivadiversityoverview}. 
The MIR objective can additionally exploit temporal diversity, enabling identification of Gaussian sources with distinguishable second-order structure. 
For real, zero-mean WSS disturbance $e_n$, defining the autocovariance function as $c_{e_n}(\tau)=\mathbb{E}\{e_n(t+\tau)e_n(t)\}$, Gaussian sources are distinguishable by second-order statistics when their autocovariance functions are nonproportional: 
\begin{equation}
    \nexists\,\alpha\neq0:\;
    c_{e_n}(\tau)=\alpha^2 c_{e_m}(\tau),\quad \forall\tau,
    \qquad n\neq m.
\end{equation}
By jointly exploiting temporal diversity and HOS, MIR broadens identifiability beyond what either source of statistical information provides alone. When the disturbance processes are identifiable up to permutation and scaling, the acyclic SEM constraints in \ref{sec:linearsem} yield the causal coefficient matrix.

\begin{figure*}[t!]
  \centering
  \includegraphics[width=1.0\textwidth]{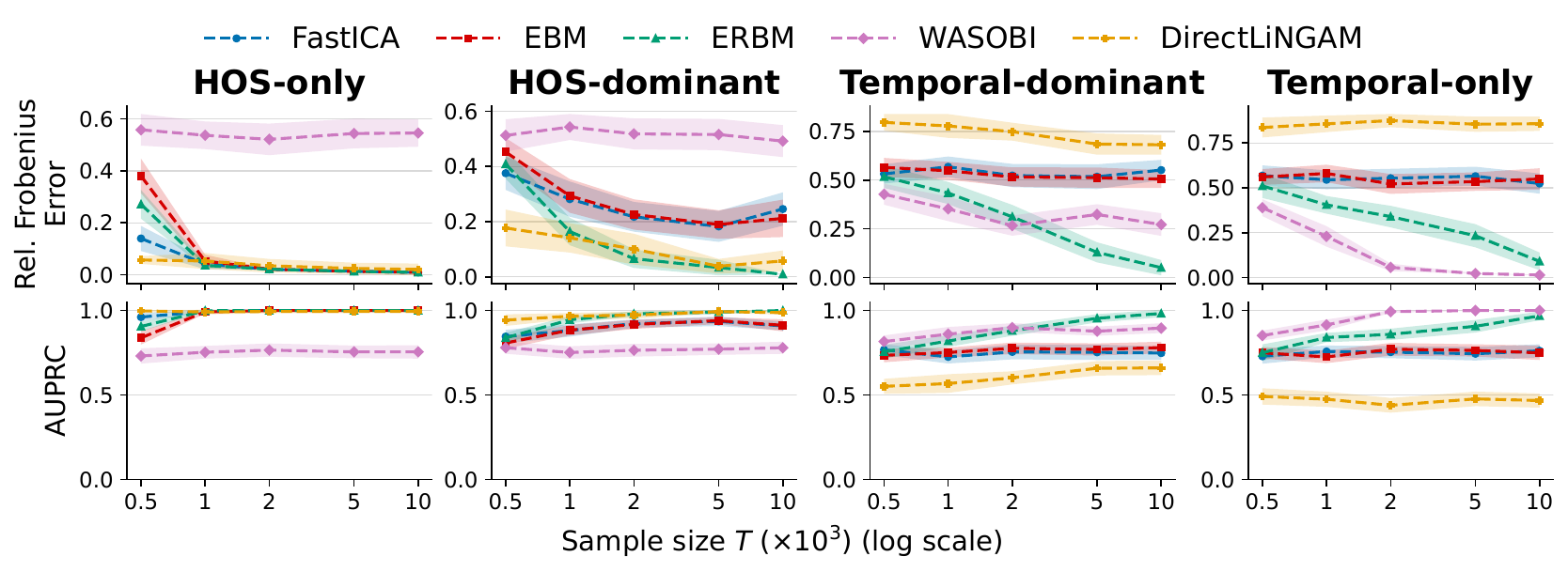}
  \vspace{-6mm}
  \caption{Performance of methods versus sample size across different experiment settings. BSS algorithm labels refer to the use of the corresponding method for SEM estimation. Curves show the mean metric values over 50 trials; shaded regions show 95\% confidence intervals. Lower is better for relative Frobenius error (top row), while higher is better for AUPRC (bottom row).}
  \label{fig:performance-grid}
\end{figure*}

\section{Experimental Results}
\subsection{Synthetic Simulations} \label{sec:sim}
We compare four BSS algorithms for linear causal structure recovery. FastICA \cite{fastica}, the algorithm employed in original ICA-LiNGAM \cite{lingam}, and entropy bound minimization (EBM) \cite{icaebm} exploit HOS only, with EBM providing a more flexible source density model. Weights-adjusted second-order blind identification (WASOBI) only uses second-order statistics, and entropy rate bound minimization (ERBM) jointly exploits both diversities. We additionally include DirectLiNGAM since it also relies on non-Gaussianity despite not using ICA.

We consider four statistical-diversity settings to compare the random-variable and random-process formulations. 
\textbf{HOS-only} uses generalized Gaussian distribution (GGD) disturbances with equal numbers of evenly spaced shape parameters \(\beta\) in \([0.2,0.8]\) and \([1.5,8.0]\), ensuring non-Gaussianity. 
\textbf{Temporal-only} uses Gaussian AR(1) disturbance processes, with AR coefficients chosen from evenly spaced grids over $[-0.9, -0.1] \cup [0.1,0.9]$ to ensure temporal diversity. 
\textbf{Temporal-dominant} uses $20\%$ GGD and $80\%$ Gaussian AR(1) disturbances, while \textbf{HOS-dominant} uses the opposite $80\%/20\%$ configuration. 

In each setting, we generate \(N=20\) variables with a random acyclic adjacency matrix \(\B\) of density \(d=0.1\), sampling edges from the \(N(N-1)/2\) permissible pairs. Nonzero coefficients are uniformly sampled from \([-1.5,-0.5]\cup[0.5,1.5]\). 
We report the relative Frobenius error, $||\B-\widehat{\B}||_F/||\B||_F$, to assess causal coefficient recovery, and AUPRC, computed as average precision, to assess edge detection.

The simulation results are shown in Fig. \ref{fig:performance-grid}. HOS-based methods perform well when non-Gaussianity is the dominant source of diversity, but degrade as Gaussian disturbances become dominant and fail to recover the true structure in the \textbf{Temporal-only} setting. Despite its more flexible density model, EBM performs similarly to FastICA in these settings, suggesting that density flexibility alone does not compensate for the omission of temporal information. WASOBI performs best in \textbf{Temporal-only}, where sample dependence is the only available diversity, but performs poorly in \textbf{HOS-only} and \textbf{HOS-dominant}, consistent with its reliance on temporal structure. Both mixed settings contain multiple Gaussian disturbances, which are not individually identifiable from marginal non-Gaussianity, and multiple disturbances with independent samples, which are indistinguishable through temporal covariance alone. Hence, neither source of diversity alone guarantees complete causal recovery in these cases. 
Settings where only a single diversity is abundant also highlight a sample-efficiency tradeoff. In Temporal-only, WASOBI reaches an AUPRC close to one around \(T=2000\), whereas ERBM approaches comparable ranking performance near \(T=10000\). Similarly, methods such as FastICA provide better performance at low sample sizes in \textbf{HOS-only}. Thus, a specialized method is effective when its assumptions perfectly match the data, while joint exploitation performs more consistently across heterogeneous disturbance settings.

\subsection{Application to Real fMRI Data}
\subsubsection{Dataset Details}
We analyze resting-state fMRI from $K=250$ healthy controls in the BSNIP cohort \cite{bsnip1,bsnip2}, with $T=224$ time points per subject and TR = 2 s. Preprocessing included motion and slice-timing correction, Montreal Neurological Institute normalization, resampling, 6-mm smoothing, and brain masking. We employ threshold-free constrained independent vector analysis (tf-cIVA) \cite{arctfiva}, using the 53 NeuroMark templates as spatial references \cite{neuromark}, to obtain aligned components across subjects. We restrict the analysis to $N=7$ intrinsic connectivity networks (ICNs), comprising five subcortical (SC) and two auditory (AU) components. We standardize each component time course within each subject and concatenate the time courses across subjects along the temporal axis to form $X_{\rm group}\in\mathbb R^{N\times KT}$. 
This places estimation in a more favorable sample-to-dimension regime while enabling assessment of both within- and cross-domain ancestral stability. 

\subsubsection{Stability Analysis}
In real fMRI data, the ground-truth causal structure is unknown, so we compare methods based on the stability of their estimated causal relationships. 
We employed subject-level bootstrap resampling \cite{bootstrapbook}, sampling \(K\) subjects with replacement and concatenating their representations along the temporal axis to form \(\X_{\rm group}\in\R^{N\times KT}\). 
Stability was quantified by comparing causal structure estimates across $B=500$ bootstrap samples. 
For each bootstrap adjacency matrix, we compute its transitive closure to determine ancestral relationships among the selected ICNs. 
We compute the probability of some variable $i$ being an ancestor of a variable $j$ based on the rate of occurrence of such relation: 
\begin{equation}
    P_{ij} = \frac{1}{B} \sum_{b=1}^{B} {\mathbf{1} \{i \in {\rm anc}(j) \text{ in bootstrap } b\}}.
\end{equation}
${\rm anc}(j)$ denotes the variables reaching $j$ through a directed path. We summarize ancestral direction stability by the mean absolute difference between opposing ancestral frequencies:
\begin{equation}
    S_{\rm anc} = \frac{2}{N(N-1)}\sum_{i < j} {|P_{ij} - P_{ji}|}.
\end{equation}
Larger values indicate more reproducible preference for one ancestral direction. Low values reflect either directional ambiguity or frequent incomparability, whose frequency is \(q_{ij}=1-P_{ij}-P_{ji}\). Incomparability was generally infrequent, so the score primarily reflects directional consistency in these data. 
\begin{figure}[t] \label{fig:anc_prob}
    \centering
    \includegraphics[width=1.00\linewidth]{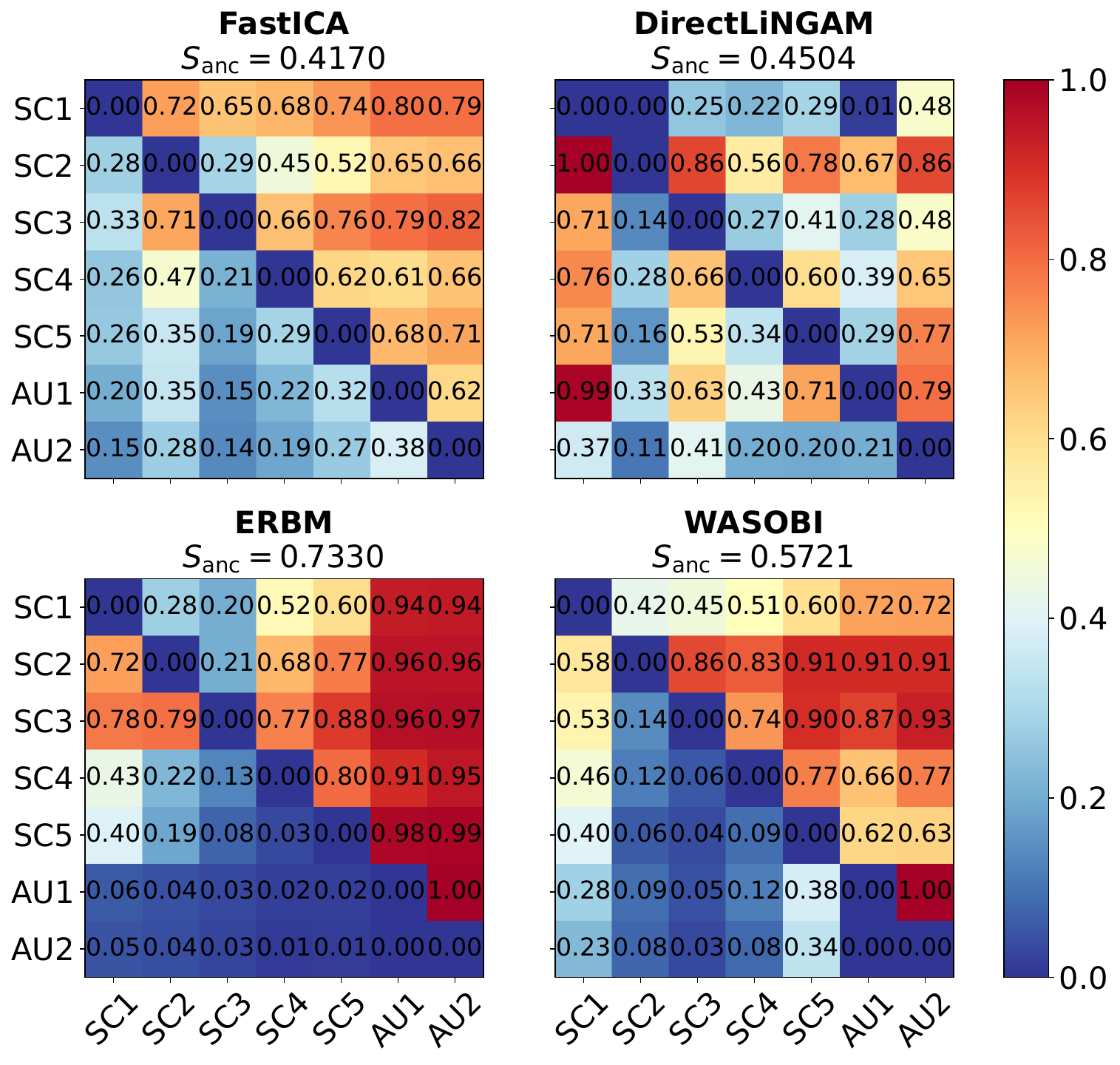}
    \caption{Bootstrap ancestral probabilities for five SC and two AU ICNs. $S_{\rm anc}$ summarizes directional ancestral stability, with higher values indicating better directional consistency.}
    \label{fig:anc_prob}
\end{figure}    

We show the bootstrap ancestral probabilities in Fig. \ref{fig:anc_prob}, with the mean ancestral stability score $S_{\rm anc}$ reported for each method. Each entry gives the fraction of bootstrap estimates in which the row ICN is an ancestor of the column ICN. Incomparable outcomes are omitted from the visualization but remain included in the bootstrap denominator. 
ERBM yields the highest directional ancestral stability (\(S_{\rm anc}=0.7330\)), followed by WASOBI (\(0.5721\)), DirectLiNGAM (\(0.4504\)), and FastICA (\(0.4170\)). 
ERBM consistently places the AU components downstream of SC, with all SC-to-AU ancestral probabilities exceeding $0.90$. 
WASOBI favors the same pattern with less uniformity. Both methods also assign AU1 as ancestor of AU2 in all estimates. 
Greater bootstrap stability does not imply agreement across methods: ERBM favors SC3 as an ancestor of SC2 (\(P_{3,2}=0.79\)), whereas WASOBI favors the opposite direction (\(P_{2,3}=0.86\)). Thus, temporal information improves directional reproducibility without necessarily establishing a unique causal graph. Causal interpretation remains limited by the assumed common group-level structure, subject-boundary artifacts from concatenation, and unobserved confounders outside the SC/AU components. 

\section{Conclusion}
We introduced a random process formulation for BSS-based estimation of linear causal structures. The proposed framework enables the joint use of sample dependence and HOS, extending identifiability beyond the non-Gaussian disturbance assumption of LiNGAM. Incorporating sample dependence enables causal structure identification with Gaussian disturbance processes when their covariance functions are sufficiently distinct. 
Synthetic experiments support the premise that jointly exploiting HOS and sample dependence improves causal structure recovery across diverse disturbance settings. HOS-based methods fail when non-Gaussianity is insufficient for identification, whereas methods relying solely on sample dependence perform poorly when the temporal diversity is limited. This contrast highlights the importance of jointly exploiting both forms of diversity. Experiments on real fMRI data further support this finding, with BSS methods jointly exploiting HOS and sample dependence producing more stable causal relationships across bootstrap samples compared to methods that rely only on HOS. These results demonstrate that exploiting statistical diversity beyond non-Gaussianity can broaden the identifiability of linear causal structure across a wider range of disturbance conditions.

\section{Acknowledgments}
This work was supported in part by grants NSF 2316420, NIH R01MH118695, NIH R01MH123610, and NIH R01AG073949. The hardware used in these studies is part of the UMBC High Performance Computing Facility (HPCF). HPCF is supported by the U.S. NSF through the MRI program (grant nos. CNS-0821258, CNS-1228778, and OAC-1726023) and the SCREMS program (grant no. DMS-0821311), with additional support from the University of Maryland, Baltimore County (UMBC). Authors would like to thank Vince D. Calhoun for helpful comments and feedback.

\bibliographystyle{IEEEbib}
\bibliography{refs}

\end{document}